\documentclass[letter,man,natbib]{apa6}

\usepackage[english]{babel}
\usepackage[utf8x]{inputenc}
\usepackage{amsmath}
\usepackage{graphicx}
\usepackage[colorinlistoftodos]{todonotes}
\usepackage{color}

\usepackage{multirow}

\title{Impact of Blockchain Technology on Electric Power Grids -- A case study in LO3 Energy}
\shorttitle{Blockchain \& LO3 Energy}
\author{Sakineh Khalili$^1$, Vahid Disfani$^2$, Mo Ahmadi$^1$}
\affiliation{\small{$^1$ Gary W. Rollins College of Business, University of Tennessee at Chattanooga, TN 37403\\$^2$ ConnectSmart Research Laboratory, Department of Electrical Engineering, University of Tennessee at Chattanooga, TN 37403}}

\abstract{The increasing amount of distributed energy resources including renewable energy systems and electric vehicles is expected to change electric power grids significantly, where conventional consumers are transformed to prosumers since they can produce electricity as well. In such an ecosystem, prosumers can start offering their excess energy to supply demands of the other customers on the grids behind the meter without interference of distribution system operators (DSO). Besides, DSOs require more accurate and more frequent data form prosumers' net demand to be able to operate their network efficiently. The main challenge in these new distribution grids is the amount of data that needs to be collected in this platform is unbelievably high, and more immortally, prosumers will likely refuse to share their information with DSOs due to their potential privacy and economic concerns. Blockchain technology as an efficient distributed solution for management of data and financial transactions, has been considered to solve this trust issue. With blockchain-based solutions, data and financial transactions between all parties will take placed through distributed ledgers without any interference from an intermediary. In this paper, impacts of blockchain technologies on electric power industry is studied. The paper specifically focuses on LO3 Energy --one of startups applying blockchain to electric power grids,-- their blockchain-based solution called Exergy, and their use cases to implement such solutions.}

\begin{document}
\maketitle

\section{Introduction}
\label{sec:introduction}

Information Technology (IT) benefits all businesses by enabling them to enhance their data storage systems, process transactions faster, disseminate information more effectively, and to improve quality of their products and services in a more efficient manner. In short, adoption of IT enables businesses to (1) solves their problems faster and more simply through computer applications, (2) make their decisions and respond to changes rapidly by studying their available data and collecting their team members' inputs, and (3) to enhance their customer service quality by connecting with their customers, hearing their stories first hand, and providing prompt assistance \citep{eason2014information}.

Among these technologies, blockchain technology emerged from its use as verification mechanism for cryptocurrencies in 2008 and heads to a broader field of economic applications \citep{mengelkamp2018blockchain}. Blockchain has been implemented in various industries including finance and insurance \citep{tapscott2017blockchain, eyal2017blockchain}, information and communication \citep{huckle2016internet, huh2017managing}, healthcare \citep{mettler2016blockchain,peterson2016blockchain}, and energy \citep{aitzhan2018security, kim2018study, mengelkamp2018blockchain, mengelkamp2018designing}. Figure \ref{fig:StatupShare} illustrates share of startups developing blockchain technology for different industry sectors \citep{friedlmaier2017disrupting}.

In energy sectors alone, around 120 energy blockchain startups raised a total of \$324 million in 2017 \citep{Muzzy2018Blockchain}. By October 2017, there were 15 international firms leading in energy blockchain from which 6 firms --Drift, GridPlus, ImpactPPA, LO3 Energy, Power Ledger, SolarCoin-- are based in the United States, 8 reside in Europe, one is located in Asia, and one in Africa. \citep{Deign2017Firms}. 

This paper focuses on LO3 Energy which is one of the US-based companies developing blockchain solutions for electric power systems. LO3 Energy has proposed a distributed ledger system, called Exergy, which functions across grid-connected hardware and advances electricity market design and technology in tandem \citep{LO3Business2017} 

\section{Blockchain Technology}
\label{sec:blocchain}
Blockchain technology was first emerged as a cryptocurrency system and the underlying technology of Bitcoin in 2008  and swiftly headed toward a broader filed of economic applications \citep{noor2018energy}. This technology is expected to revolutionize the functioning of transaction systems and enable fully distributed market platforms \citep{yuan2016towards}.

A blockchain system consists of distributed ledgers, decentralized consensus mechanisms, and cryptographic security measures. Blockchain technology provides a transparent and valid record of past transactions which may not be changed retrospectively and allows faster and more efficient resolutions of conflicts and dismantles information asymmetries without any need for governing intermediaries \citep{beck2016blockchain}. In other words, with blockchain, contracts are embedded in digital codes and stored in shared, transparent databases protected from deletion, tampering, and revision. Using blockchain, all agreements, processes, tasks, and payments will have digital records and signatures which can be identified, validated, stored, and shared. Thus, intermediaries like lawyers, brokers, and bankers might no longer be needed for verification process or conflict resolutions \citep{iansiti2017truth}.

\subsection{How Blockchain Works}
Blockchain is a chain of blocks that contains information and includes a distributed ledger which is completely open to everyone. Each block contains some data, the hash (address) of the block, and the hash of the previous block.  The data stored inside a block depends on the type of blockchain. The hash can be compared with a fingerprint, is unique, and identifies the block and all of its contents. The third element inside each block is the hash of the previous block to create the chain. This effectively creates a chain of blocks and it is this technique that makes a blockchain very secure. There is one more way that blockchains secure themselves and that is by being distributed. Instead of being a central entity to manage the chain, blockchains use the peer-to-peer network where everyone is allowed to join. Anyone who joins the network, they get the full copy of the blockchain. When someone creates a new block, that new block is sent to everyone on the network, and each node verifies validity of the block. If all the nodes in this network verifies the validity, consensus is created and each node adds the block to their own blockchain. Blocks that are not valid will be rejected by other nodes in the network. If hackers wants to corrupt the database, they need to hack all blocks on the chain, redo hash calculations for each block, and take control of more than 50\% of the peer-to-peer network. Since it is s very unlikely action, blockchain is considered extremely secure.\citep{singhal2018blockchain}

\subsection{Blockchain Principles}
Blockchain transactions are continuously verified, cleared, and stored by the network in digital blocks connected to previous blocks, thereby creating a chain. The structure permanently time-stamps and stores exchanges of value. Since each block must refer to the preceding block to be valid, altering the ledger is almost impossible. To steal anything of value, a thief would have to rewrite its entire history on the blockchain. Collective self-interest ensures the blockchain’s safety and reliability. Therefore, we think blockchain provides a powerful mechanism for blowing traditional and centralized models, such as that of the corporation, to bits.

Blockchain technology is constructed by the following 5 underlying, basic principles \citep{iansiti2017truth}:

\begin{enumerate}
    \item \textbf{\textit{Distributed Database:}} There is no single party in the blockchain controlling the data or the information. Instead, all parties have access to the entire database and its complete history and are able to verify the records of their transaction partners directly, without any intermediary. 
    
    \item \textbf{\textit{Peer-to-Peer Transmission:}} Blockchain is based on direct data communication between parties (peers) instead of through a central node. While receiving a new block of data, each party stores after validation and forwards it to all other parties.
    
    \item \textbf{\textit{Transparency with Pseudonymity:}} In blockchain, all transactions and their associated values are transparent to  all parties with access to the system. On the other hand, parties on a blockchain are identified by a unique 30-plus-character alphanumeric addresses, and transactions occur between blockchain addresses. Parties may opt to remain anonymous or provide proof of their identity to other parties. 
    
    \item \textbf{\textit{Irreversibility of Records:}}
    In blockchain, each block of data is identified by an address called hash number and is linked to its previous data block in blockchain by containing its hash number. Various computational algorithms and approaches are deployed to ensure that the recording on the database is permanent, chronologically ordered, and available to all others on the network. Once a transaction enters the database, the records cannot be changed since they are linked to its previous block, and corruption of one block invalidates all down-stream blocks. 
    \item \textbf{\textit{Computational Logic:}}
    Blockchain transactions can be tied to computational logic and in essence programmed. The digital nature of the ledgers enables users to set up algorithms and rules that automatically trigger transactions between parties.

\end{enumerate}

In short, blockchain technology offers three main advantages \citep{tapscott2017blockchain}: 
\begin{itemize}
    \item It is \textbf{\textit{distributed}} and runs on computers provided by volunteers around the world, so there is no central database to hack.
    \item It is \textbf{\textit{public}}, and anyone can view it at any time because it resides on the network;
    \item It is \textbf{\textit{encrypted}} and uses heavy-duty encryption to maintain security.
\end{itemize}

\subsection{Need for Blockchain in Electric Power Grids}
In the US, there is an 86\% inefficiency in converting energy and delivering it to the end users as the useful product, described by physicists as \textbf{\textit{exergy}}, which creates costly health and environmental challenges as well as significant loss of economic values. Such problems will persist as long as utilities use outdated constructs of its consumers as rate payers and miss the opportunity and value only a customer or end user can provide \citep{LO3Business2017}.

Relatively high price of electricity, global commitments to halting climate change, a slew of energy efficiency policies, and subsidies for renewable power have motivated a new wave of technological innovation such as smart energy services, battery storage, and rooftop solar Photovoltaic (PV) systems. These technologies making it possible for customers to not only save energy, but to produce their own power in factories and homes and become \textbf{\textit{prosumers}}, consumers with capability of producing energy. Therefore, for optimal operation of their power grids, utilities need to have visibility into millions of new consumer devices and distributed energy resources popping up at the grid edge. 

However, end users have no obligation to reveal their personal information and may refuse to share their energy consumption and production data with utilities due to various concerns such as privacy and economic interests. Also, aggregation of all data from all customers in utilities' databases is not a secure solution. Thus, Blockchain as a reliable distributed data management system has been investigated by several firms in electric power sectors, including LO3 Energy.

Employing Blockchain technology enables utilities to have access to customers' data for optimal asset management as well as to benefit from their new roles as new value domains which customers will provide, including grid management services --capacity, real power, reactive power, and frequency regulations \citep{LO3Tech2017}. Such data access allows this new values to be unlocked in future structure of power grids. This data not only includes customers' energy demand or consumption, but also accounts for the state of the grid, the time and location of production, and consumption requirements of a multi-party electricity system \citep{LO3Tech2017}. 

Other than data management, adoption of blockchain in transactions between customers and utilities as well as between customers themselves removes friction in buying and selling real value-added energy services. Thus, smart contracts and tokenization are seriously considered in application of blockchain in power systems \citep{kim2018study}.

\subsection{Implementation of Blockchain by LO3 Energy}

LO3 ENERGY has proposed and implemented blockchain by developing the concept of Exergy consisting of a distributed ledger system which functions across grid-connected hardware, a token system for transactive energy, and a foundation which advances market design and technology in tandem. Exergy is based on a market model called transactive energy, which refers to economic and control techniques that enable broad participation in the new multi-party energy system, and designed to provide all market participants more opportunities to connect customers to energy system value. Exergy enables customers and utilities to buy and sell value by ensuring data availability for transactions at all electricity market levels. The Exergy blockchain is the private distributed ledger which secures and manages the data needed to create marketplace transactions, and settles those transactions cost-effectively.  Thus, blockchain is implemented by LO3 Energy, through Exergy, in order to manage both data and transactions. \citep{LO3Tech2017,LO3Business2017,LO3ExeSum2017}

\subsubsection{Exergy Data Management System}
For information systems, Exergy develops distributed ledger accounts to enable data within the energy system to be securely transacted for different marketplace behaviors such as buying or selling, and setting and reacting to prices. Exergy runs on a private, permissioned blockchain --where certain actions are just permitted by certain players within blockchain-- through a network of globally distributed computing nodes. Some of these Exergy-enabled devices, which are developed by LO3 Energy as well as other 3rd party providers, control, manage and validate actual power flows on the electricity grid. Others simply verify and monitor marketplace rules and activity. 

An Exergy-based data Collection system collects, controls and secures the data to provide a wide range of advantages throughout power markets and ecosystems, including 
(1) efficient and adaptive market pricing,
(2) improved system reliability and flexibility Pathway for technological innovation,
(3) data needed to develop additional direct and derivative markets,
(4) improved balance of risk and reward for asset owners, and
(5) a rich, interactive future for an energy industry serving informed communities \citep{LO3Business2017}.

\subsubsection{Exergy Transaction Management System}
For transaction management systems, Exergy System’s top layer, the “Exergy Layer,” will use a blockchain to establish and manage a global network of power market participants. It also establishes location, security and proof of ownership. The Exergy token (“XRG”) is an ERC20-compliant exchange medium and will be used to attract consumers, prosumers and communities to the marketplace. Consumers and prosumers will be rewarded for their participation in local energy markets, and other market participants such as generators, aggregators and service providers can also acquire and hold tokens. The XRG token is staked to access the marketplace, where available data and transaction behavior can be verified into blocks and, where necessary, linked to actual control of assets and settlement of energy transactions.

An Exergy-based transactive energy system (1) allows market participants around the world to buy and sell electricity and (2) offer grid services based on what the new decentralized system needs to operate at maximum efficiency, reliability and flexibility. The Exergy token (XRG) also provides the incentive for energy prosumers, consumers, and commercial market players to participate in an energy marketplace where consumers are front and center. 

\subsection{LO3 Energy Use Cases for Blockchain Application}

Through blockchain technology and innovative solutions, LO3 Energy has developed Exergy, a permissioned data platform to create localized energy marketplaces for transacting energy across existing grid infrastructure. According to their white-papers \citep{LO3Tech2017,LO3Business2017,LO3ExeSum2017}, LO3 Energy team have identified 4 use cases where blockchain technology has been or is going to be implemented within its organization. These use cases are detailed below and summarized in Table \ref{table:usecases} on page \pageref{table:usecases}.

\subsubsection{Use Case 1 -- Peer-to-Peer Energy: Brooklyn Microgrid}

LO3 Energy has already implemented a peer-to-peer use case, the Brooklyn Microgrid, to explore full market value of peer-to-peer energy. Blockchain technology is used in this platform to let prosumers and consumers exchange their data of excess energy and their power demand with each other without any intermediary managing their communications. This data exchange allows prosumers to transact their excess energy autonomously in real-time with consumers on the platform in their local marketplace. Blockchain technology is then used to clear value transactions between supply and demand parties using XRG tokens.

\subsubsection{Use Case 2: Microgrid}
A microgrid is an ecosystem of connected prosumer and consumer energy assets, where energy is generated, stored, and transacted locally and enhances efficiency, resiliency and sustainability within communities. Applying Exergy offers additional value to microgrid project developers, operators and other hardware vendors. System-wide operational efficiency is improved by realizing responsive load management and improving return on investment (ROI) and economics for developers and operators. The blockchain-based Exergy (1) allows monitoring of excess capacity and potential to sell to other microgrids, and (2) finds value for the storage and other services available from the microgrid. Data transparency and uniformity across project portfolios --as a result of adopting blockchain technology-- will further support future commercial investments in microgrid ecosystem.
Consumers and prosumers’ roles in the microgrid use case look to similar to Use Case 1. The difference is that in Use Case 2, there are several distinct microgrids connected to one another over the existing distribution network, and the distribution system operators (DSO) can open up their network to let electricity flow between microgrids.

\subsubsection{Use Case 3: Distribution System Operator (DSO) Use Case}
With the current transformation in customers' roles in power grids and their capability of generating energy, the distribution network operator role is also evolving as more activity is required by them to ensure real and reactive power requirements are met in a world of distributed energy resources. Microgrids and individual customers are also able to offer different ancillary services including spinning reserve, frequency regulations, voltage control, and demand response. Through blockchain, the distributed system operator is granted access to consumer data like building management systems. Using price as a proxy, the DSO manages energy use, load balancing, and demand response at negotiated rates. Blockchain-based Exergy solution are developed for DSO and microgrids to securely share their supply and demand data with each other and coordinate exchange of ancillary services and associated value transactions. 

\subsubsection{Use Case 4: Electric Vehicle smart charging}
Number of electric vehicle (EV) on roads is exponentially increasing every year, so their adoption is accelerating rapidly. In 2016 the number of EVs reached one million for the first time; In ealry 2018, there were more than 2 million EVs, and their number is forecasted to be 40 to 70 million in 2025. With an average power consumption of 3 MWh per EV per year, these new cars will consume the output of 100 new 400-MW power plants every year. It will be very important for grid management where, when, and how drivers charge their EVs . Today, in the US alone, that market is \$250 million. Globally, by 2025, it will reach \$25 billion. When a charging station --public or private-- or an electric vehicle has a surplus of energy, it is made available for purchase on the local network. Consumers can set budgets and be alerted to the availability via mobile app. 

Blockchain-based solutions such as Exergy not only enables EVSEs (EV supply equipment) --also known as EV charging stations-- and EVs to share their parking availability and their charging demand with each other to be matched more efficiently, but also it enables DSOs to be aware of where charging events will take place and how they would affect the operation of the distribution system. All data and value transactions can be securely taken placed through blockchain without any need for intermediaries.

\section{Exergy Competitive Advantages in Electric Power Grids}
Employing blockchain in electric power grids is proved to be very \textbf{\textit{valuable}}. As detailed above, Exergy is a blockchain-based solution developed by LO3 Energy to realize data and value transactions between supply and demand in electric power grids. In all 4 use cases which LO3 Energy team consider to expand their business, Exergy is used as a solution to build trust between demand and supply parties to securely share their availability and demand data of different services without any need for regulating intermediaries. Blockchain technology is also used to record value transactions between these parties as an exchange for service provided by supply to demand side. Exergy uses XRG tokens to monetize these transactions. Through these solutions, LO3 Energy has offered several technical advantages to electric power grids in different levels. 

Further, adoption of blockchain in operation of electric power grids is also \textbf{\textit{rare}}. There are few firms actively working in this business. Use cases defined by these firms are rather new and have not been completely implemented yet. LO3 Energy team has already implemented one of their use cases in Brooklyn Microgrid project in new York, as the first peer-to-peer trading pilot project in history, which keeps them some steps ahead of the other firms competing with them. 

The complexity of the blockchin technology makes its adoption in electric power systems \textbf{\textit{hard to imitate}}. Blockchain technology involves a very complicated mathematical background and data management algorithms which requires high level of knowledge and expertise. Thus it is very difficult for new firms to enter this market without investing time, energy, and assets. 

The main competitors of startups like LO3 Energy are distribution system operators (DSOs). However, current operation procedures adopted by DSOs are well established and are used for decades, and development and application of blockchain is \textbf{\textit{hard to substitute}} by them because (1) utilities are normally reluctant to significant changes in their operation procedures until the modifications are well proved and tested and (2) blockchain technology will likely threaten or even end their authorities over their network and customers' data. This fact, on the other hand, makes it difficult and costly for startups like LO3 Energy to adopt blockchain technologies when it comes to the use cases 2-3 where DSO is one of the involved parties.
 
\section{Conclusion}
In this report, we have studied impacts of blockchain technologies on electric power grids, with a particular focus on LO3 Energy as one of the firms developing blockchain solutions for energy sector. LO3 Energy has developed a technology called Exergy to implement blockchain in distribution power grids. Exergy includes both data and financial transaction management systems via blockchain for several use cases including peer-to-peer energy sharing, microgrid management, DSO services, and EV charging management. They have one active project in Brooklyn, New York, where prosumers who own rooftop solar PVs share their excess energy with other consumers in the grid. Benefits of Exergy in these use cases are comprehensively discussed in the report. 

There are two major concerns associated with blockchain solutions in power industry which lead to the following two recommendations:

\begin{enumerate}
    \item Although distributed nature of blockchain makes it extremely secure, it also mandates all parties in the network to invest on creating database on their nodes. That is, the cost associated with database is now multiplied by the number of parties, which makes this technology extremely costly. Right now, in pilot projects like Brooklyn Microgrid, the solution looks viable because of the limited number of active prosumers, but its adoption to larger grids with higher number of parties would be very expensive. Thus, a major improvements in technology of computer memories is needed to reduce implementation costs of such distributed solutions.
    
    \item Consumers and prosumers sound very interested in adopting these technologies in power industry since they resolve their privacy and economic concerns very well. On the other hand, such technologies significantly reduce the authority of the utilities or DSOs on distribution power grids. In addition, to maintain the minimum reliability needed in distribution systems, DSOs always want to make sure new solutions are completely compatible with their current procedures and technologies before they start adopting them. These concerns make DSOs reluctant against these evaluations on the grid. Startups such as LO3 Energy need to engage DSOs and establish a full cooperation with them in their technology development process. When DSOs are parts of the business, the new technologies will be more effective and practical.
    \end{enumerate}

% Commands to include a figure:
\begin{figure}
\centering
\resizebox{\textwidth}{!}{
\includegraphics[width=0.9\textwidth]{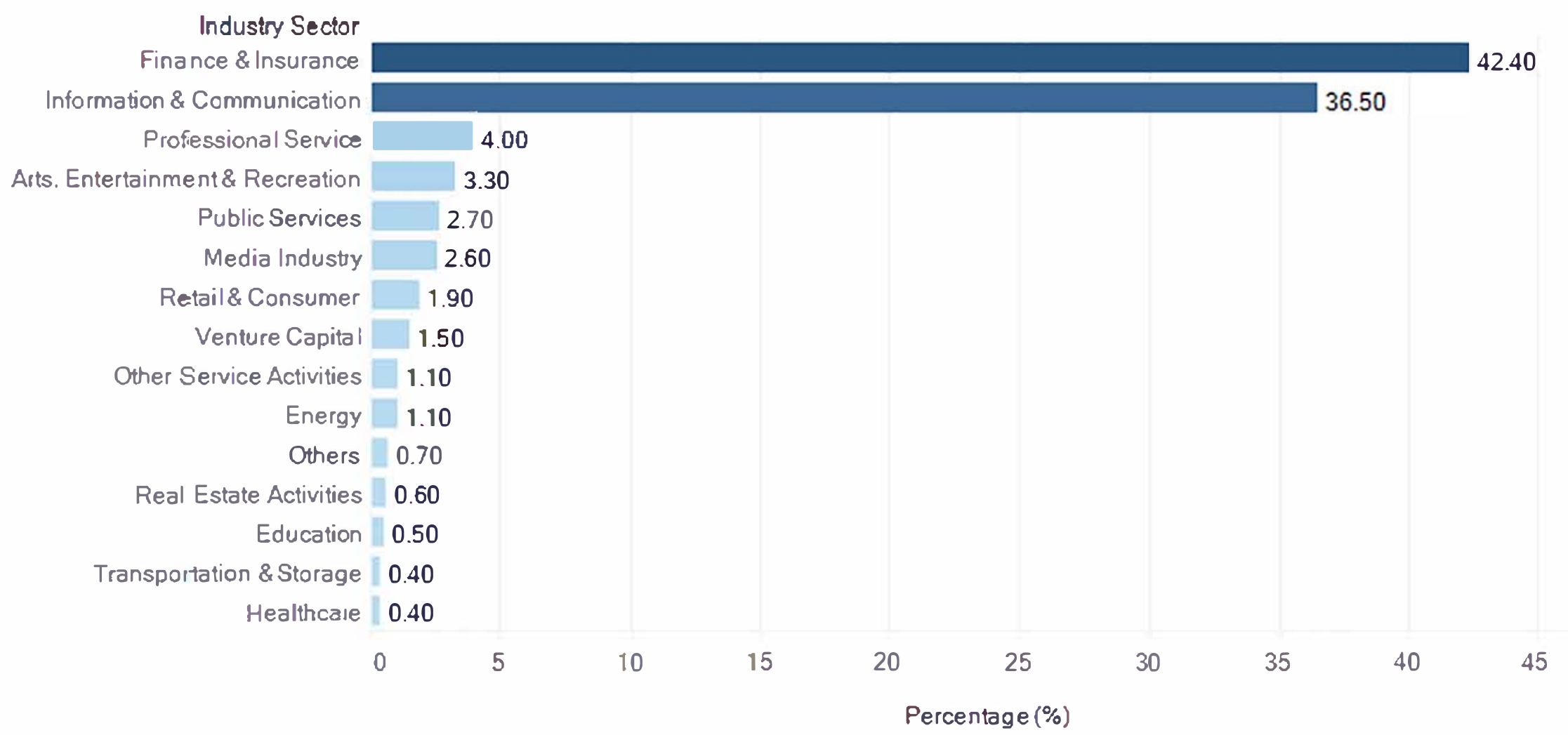}}
\caption{Percentage of blockchain startups operating in each industry sector.}
\label{fig:StatupShare}
\vspace*{\floatsep}% https://tex.stackexchange.com/q/26521/5764
\centering
\resizebox{\textwidth}{!}{
\includegraphics[width=0.9\textwidth]{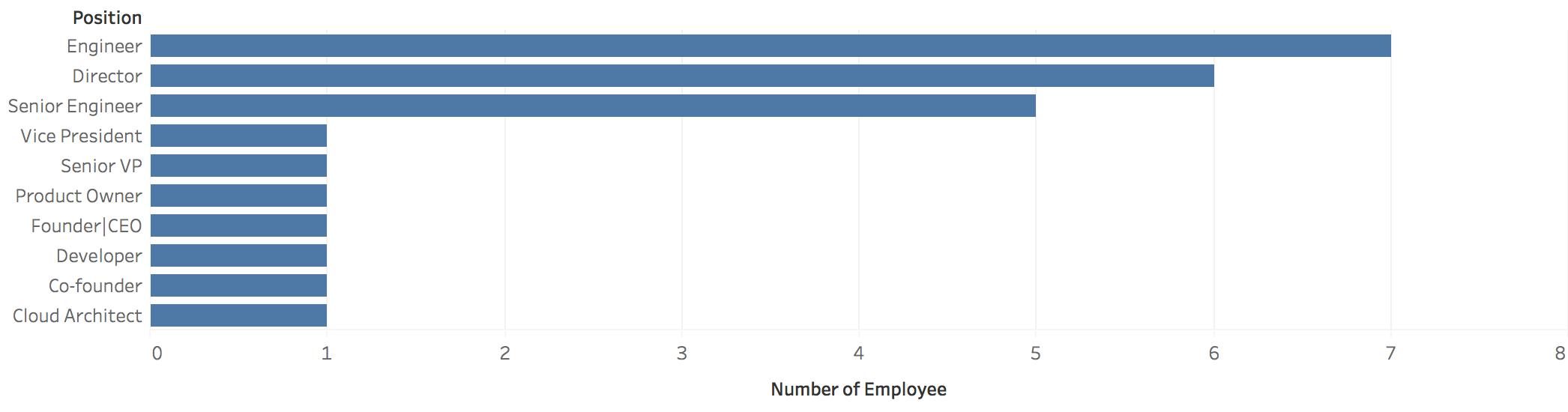}}
\caption{Number of Employee currently working in LO3 Energy.}
\label{fig:LO3Employee}
\end{figure}

\begin{table}
\resizebox{\textwidth}{!}{
  \begin{tabular}{|p{2.5cm}|p{2.5cm}|p{5cm}|p{5cm}|p{5cm}|p{5cm}|}
    \hline
      \multicolumn{2}{|p{5cm}|}{Parameters/Use Case} &
      1: Peer-to-Peer Energy &
      2: Microgrid &
      3: Ancillary Services for DSOs, e.g. EPB&
      4: Electric Vehicle Smart Charging\\
    \hline
    \hline
    
    \multirow{2}{*}{Parties} &
    Supply & 
    Prosumer & Microgrid 1 & 
    All Microgrids & 
    EVSEs (EV Chargers)\\
      \cline{2-6}
    & 
    Demand & 
    Consumers & 
    Microgrid 2 & 
    DSO & 
    EV \\
    \hline 
    
    \multirow{3}{2.5cm}{Translated Data via \textbf{Blockchain}} &
    Supply & 
    Excess Energy/Price & 
    Excess Energy/Price & 
    Availability to Provide Service + Service Price & 
    Parking Availability and Maximum Power Rate + Desired Price\\
    \cline{2-6} 
    & 
    Demand & 
    Power Demand/Budget& 
    Power Demand/Budget & 
    Need for Ancillary Services + Budget & 
    Charging/Parking Demand + Budget\\
    
    \cline{2-6}
    & Other & 
    N/A &
    Network Constraint by DSO to make sure transaction is feasible & 
    Network Constraint by DSO to make sure transaction is feasible & 
    N/A
    \\
    \hline   
    
    \multicolumn{2}{|p{5cm}|}{Transacted Service through \textbf{Blockchain}}&
    Excess energy of prosumer supplies consumers' demand & 
    Excess energy from MG1 to MG2 demand. DSO checks feasibility of power exchange & 
    Ancillary services INCLUDING reserve, frequency and voltage regulation, demand side management, and demand response  & 
    EV Charging/Parking\\
    \hline  
    
    \multicolumn{2}{|p{5cm}|}{Transacted Value through \textbf{Blockchain}}&
    XRG & 
    XRG & 
    XRG & 
    XRG\\ 
    \hline
     
%    \multicolumn{2}{|p{5cm}|}{Expected Benefits}&
%    Prosumer& 
%    Microgrid 1 & 
%    All Microgrids & 
%    EVSEs\\ 
%    \hline   

%    \multicolumn{2}{|p{5cm}|}{Examples of Real Projects}&
%    Brooklyn Microgrid & 
%    Microgrid 1 & 
%    All Microgrids & 
%    EVSEs\\ 
%    \hline 
 \end{tabular}
}
 \caption{Summary of Use Cases of Blockchain-based Exergy, Developed by LO3 Energy.}
\label{table:usecases}
\end{table}

\bibliography{example}
\newpage
\section{Attachment: LO3 Energy Background}
LO3 Energy, founded 2012 in Brooklyn, NY, is developing blockchain based innovations to revolutionize how energy can be generated, stored, bought, sold and used, all at the local level. Over the last two years, the LO3 Energy team has delivered the Brooklyn Microgrid and the world’s first ever energy blockchain transaction in April 2016. The team consists of a founder and CEO, a co-founder, a senior VP, a vice president, a product owner, 6 directors, a developer, 5 senior engineer, 7 engineers, and a cloud architect Figure \ref{fig:LO3Employee}. The team has developed a unique knowledge base of integrating blockchain with physical energy generation and management assets in a regulated environment.Through blockchain technology and their own innovative solutions, they have developed Exergy, a permissioned data platform that creates localized energy marketplaces for transacting energy across existing grid infrastructure. they are just beginning to uncover the potential of the Exergy platform to influence the energy model of the future, and already the possibilities seem endless.

\end{document}